# Noninvasive measurements of gas exchange in a three-dimensional fluidized bed using hyperpolarized $^{129}$Xe NMR

Tina Pavlin[a], Ruopeng Wang[a,b], Ryan McGorty[a,c], Matthew S. Rosen[a], David G. Cory[b],
Donald Candela[d], Ross W. Mair[a,b] and Ronald L. Walsworth[a,c]

[a] Harvard-Smithsonian Center for Astrophysics, Cambridge, MA 02138
[b] Department of Nuclear Science and Engineering, Massachusetts Institute of Technology, Cambridge, MA 02139
[c] Department of Physics, Harvard University, Cambridge, MA 02138
[d] Department of Physics, University of Massachusetts, Amherst, MA 01003

**Corresponding Author:**
Ross Mair
Harvard Smithsonian Center for Astrophysics
60 Garden St, MS 59
Cambridge, MA, 02138
USA
Phone: 1-617-495 7218
Fax: 1-617-496 7690
Email: rmair@cfa.harvard.edu

**Abstract**

We present a novel NMR technique that provides a non-invasive, direct measurement of gas exchange in a three-dimensional gas-fluidized bed of solid particles. The NMR spectrum of hyperpolarized $^{129}$Xe gas in an $Al_2O_3$ particle bed displays three resolved peaks corresponding to xenon in bubbles, the interstitial spaces (emulsion), and adsorbed on particles. Modified NMR exchange and saturation-recovery sequences, together with data analysis based on an exchange-coupled set of Bloch equations, yield gas exchange rate constants between the emulsion and adsorbed phases, and between the bubble and emulsion phases. The results are in approximate agreement previously unverified predictions from well-known models of fluidized bed behavior. Incorporation of NMR imaging methodologies would straightforwardly allow similar measurements on a spatially-resolved basis.

# 1    Introduction

Granular media are encountered in a wide variety of industries. A common dynamic process that granular media undergo is gas fluidization, where solid particles are suspended in a stream of upward flowing gas [1,2], and where at higher gas flow rates, bubbles emerge from the bottom of the bed and expand while rising [2, 3]. Bubbles agitate the bed, but also provide a route for gas to escape the bed without contacting the solid particles [3]. Gas atoms can migrate from bubbles to the emulsion phase (interstitial space between closely-packed solid particles), and additionally from the emulsion to the micro-porous surface of the particles [4]. Exchange rates for both these processes have a significant impact on the effectiveness of the gas-fluidized bed.

Despite their wide application, our understanding of the dynamics of the particles and fluidizing gas is far from complete. Such systems are difficult to model numerically due to the large number of degrees of freedom and inelastic collisions among the particles [5]. Additionally, three-dimensional gas-fluidized beds are opaque, presenting difficulties when using optical techniques to probe bed behavior below the surface [6]. Previous studies have shown the utility of NMR as a non-invasive tool for studying the dynamics of the solid particles in granular systems [7-15]. However, while the $^1$H spins in certain particles give a high NMR signal they convey no direct information about the gas flow. Additionally, only the most recent of these studies focus on gas fluidized beds [12-15]. We have previously reported a preliminary attempt to observe gas dynamics in a gas-fluidized bed using NMR spectroscopy [13].

In this paper, we present a novel NMR method to non-invasively measure the gas exchange rates between bubble, emulsion and adsorbed phases in a three-dimensional gas-fluidized bed. A fluidized bed apparatus that operates in a standard vertical bore NMR magnet allows us to probe a bed of alumina particles fluidized by hyperpolarized xenon. Exchange data are acquired with variants of the traditional three-pulse exchange spectroscopy sequence [16] and the saturation recovery technique [17]. Fitting the NMR data to an exchange-coupled set of Bloch equations yields both bubble-emulsion and emulsion-adsorbed phase gas exchange rates. Although the method, as presented, was implemented using the non-spatially resolved NMR signal, incorporation of imaging methodology in the future is straightforward.

# 2    Background and Motivation

## 2.1    *Gas-exchange in fluidized beds*

A fluidized bed is comprised of a vertical column that contains the gas and solid particles, and a porous gas distributor, or diffuser, which supports the particles from below and ensures the incoming gas flow is homogeneous in cross-section. The solid particles can be classified, based on their size, density and resultant fluidizing behavior, into four categories using the Geldart Classification [18]. In this study, we use Geldart Group B particles, which have diameters between 40 and 500 µm and a moderately high density. Homogeneous fluidization without bubble formation is not generally observed for Group B particles. Instead bubbles emerge almost immediately when the bed is fluidized – a state known as bubbling fluidization. The gas in a bubbling fluidized bed may be in the bubble, emulsion, or adsorbed phase. Bubbles provide a low resistance gas flow path, and therefore expand by taking up gas from their surrounds while rising up the column [19, 20]. The emulsion phase contains two sub-phases: dense emulsion and clouds. The dense emulsion is gas surrounded by closely packed particles, while the cloud phase is gas circulating around and close to the bubble, and is generated when the rising velocity of the bubbles is greater than the upward gas velocity in the emulsion. A schematic diagram of the bubble and cloud is given in Figure 1a, while Figure 1b shows a representation of the gas-exchange processes between the gas phases described above in a bubbling fluidized bed.

Mass exchange between the gas phases described above occurs via the combined effects of gas diffusion, coherent gas flow, and particle motion carrying adsorbed gas atoms [4, 21-23]. Previous measurements of bubble-emulsion exchange in three-dimensional fluidized-beds generally used invasive procedures, such as the injection of a pocket of gas of a different species from that in the emulsion phase, followed by measurements of the depletion rate of the gas concentration in the injected bubble [19, 23-26]. The invasive probe obstructs the flow path of both the gas and particles, and inevitably disturbed the bed operation, while the injected bubbles differ significantly from naturally-generated bubbles in terms of size, density, and dynamics. Perhaps most significantly, most measurements relied on a single-bubble analysis, and did not account for the effect of multiple bubbles, possibly interacting or coalescing, on the exchange rate. Similarly, previous efforts to measure emulsion-adsorbed phase exchange were also invasive, generally introducing particles containing a removable tracer material into a fluidized bed [27-30]. The gas-solid exchange rate was then measured by observing the tracer concentrations in the gas exiting the column. These methods are extremely time-intensive, invasive, and unable to provide a spatial measure of tracer concentration along the bed. Finally, we are not aware of any reported work that measures the bubble-emulsion and emulsion-solid exchange rates simultaneously.

NMR and MRI have demonstrated their power in recent years as a tool for the non-invasive study of complex three-dimensional systems in biomedicine and materials science [31]. In contrast to methods used previously to probe fluidized beds [23-30], NMR combined with hyperpolarized $^{129}$Xe or $^{3}$He gas [32] has the potential to function as a noninvasive probe of both particle and gas dynamics in a wide range of three-dimensional fluidized beds. For most particles, the physical and chemical properties will be unchanged by the strong magnetic field used in NMR. Moreover, NMR the technique relies on naturally-formed bubbles of hyperpolarized noble gas generated at the bottom of the bed with a standard diffuser, which better represents a typical fluidized bed operation than large (> 1 cm) artificial bubbles that result from tracer gas injection. Finally, since the NMR signal represents a volume average from a region within the RF coil, measurements reflect the true, multi-bubble nature of the bed, including the effects of bubble coalescence and splitting on the observed exchange rate. This study was designed to capitalize on these benefits in order to probe accessible timescales and kinetic processes, although the geometry and chemical composition of the system may deviate from those commonly encountered in industrial applications.

*2.2  NMR Model for Fluidized Bed Study*

Our NMR model for the study of inter-phase exchange in fluidized beds is based on the classical relationship for chemical exchange between two regions. The mass transfer of a material *A* from a bubble of volume $V_b$ into the surrounding emulsion phase can be written as [21,33]:

$$-\frac{1}{V_b}\frac{dN_A}{dt} = K_{be}(C_{Ab} - C_{Ae}), \qquad (1)$$

where $N_A$ is the number of atoms of *A* in the bubble, $C_{Ab}$ and $C_{Ae}$ are the concentrations of *A* in the bubble and emulsion respectively, and $K_{be}$ is the bubble-emulsion exchange rate.

In our study, the gas-exchange dynamics may be described through the temporal and spatial dependence of the longitudinal spin magnetization, since the spin magnetization is proportional to the concentration of the species. The spin magnetization is manipulated with RF pulses and placed in an initial state that can then reveal a time-dependent variation during steady-state operation of the bed. Gas exchange between phases can therefore be incorporated into the Bloch equations in a manner similar to that derived for the chemical exchange between nuclear species [34,35]:

$$\begin{aligned}
\frac{\partial M_b}{\partial t} + u_b \frac{\partial M_b}{\partial z} &= -\frac{M_b - M_b^0}{T_1^b} - K_{be} M_b + K_{eb} M_e \\
\frac{\partial M_e}{\partial t} + u_e \frac{\partial M_e}{\partial z} &= -\frac{M_e - M_e^0}{T_1^e} - K_{eb} M_e + K_{be} M_b - K_{ea} M_e + K_{ae} M_a . \\
\frac{\partial M_a}{\partial t} &= -\frac{M_a - M_a^0}{T_1^a} - K_{ea} M_a + K_{ae} M_e
\end{aligned} \qquad (2)$$

$M_p$, $M_p^0$, $u_p$ and $T_1^p$ are the spin magnetization, spin magnetization at thermal equilibrium, upward (*z*) component of the velocity, and $^{129}$Xe longitudinal relaxation time in each phase *p* respectively, where the $p = a, b, e$ for adsorbed, bubble or emulsion phases. $K_{be}$ and $K_{eb}$ are the bubble-emulsion and emulsion-bubble exchange rates while $K_{ea}$ and $K_{ae}$ are the emulsion-adsorption and adsorption-emulsion exchange rates, all with units of $s^{-1}$. The equations above combine the effects of spin relaxation and gas-exchange on the spin magnetization in a fluid element along the gas flow. Note that the gas velocity in the adsorbed phase is zero since there is no net particle flow in the steady state.

The exchange rates, $K_{be}, K_{eb}, K_{ea}, K_{ae}$, in Eq. 2 are proportional to the surface-area-to-volume ratio of each phase, and the physical, geometry-independent, mass transfer coefficient. Since the surface through which the gas is exchanging is the same for bubble-emulsion and emulsion-bubble exchange, we can write $K_{be}V_b = K_{eb}V_e$, so that $K_{eb} = (\psi_b/\psi_e)K_{be}$. $\psi_b$ is the fractional volume of the bubble, $V_b/V$, and $\psi_e$ is the fractional volume of the emulsion phase, $V_e/V$, assuming a constant xenon number density in both phases. Similarly, $K_{ae} = (\psi_e/\psi_a)K_{ea}$, where $\psi_a$ is the effective volume of the adsorbed phase that would result from a common xenon number density in all three phases. Eq. 2 then becomes:

$$\frac{\partial M_b}{\partial t} + u_b \frac{\partial M_b}{\partial z} = -\frac{M_b - M_b^0}{T_1^b} - K_{be}\left(M_b - \frac{\psi_b}{\psi_e}M_e\right)$$

$$\frac{\partial M_e}{\partial t} + u_e \frac{\partial M_e}{\partial z} = -\frac{M_e - M_e^0}{T_1^e} - K_{be}\left(\frac{\psi_b}{\psi_e}M_e - M_b\right) - K_{ea}\left(M_e - \frac{\psi_e}{\psi_a}M_a\right). \qquad (3)$$

$$\frac{\partial M_a}{\partial t} = -\frac{M_a - M_a^0}{T_1^a} - K_{ea}\left(\frac{\psi_e}{\psi_a}M_a - M_e\right)$$

We can make estimations of the size of $K_{be}$ and $K_{ea}$. For bubble-emulsion exchange, we can use the Davidson model of gas-exchange [4] in which gas exchange occurs via diffusion and gas through-flow. For a bubble of diameter $d_b = 1$ mm, gas diffusion coefficient $D = 3$ mm$^2 \cdot$s$^{-1}$, and gas velocity in the emulsion phase $u_e = 0.5$ cm$\cdot$s$^{-1}$, the bubble-emulsion exchange rate would be ~ 100 s$^{-1}$. Conversely, the exchange between the microporous gas adsorbed on the surface of alumina particles and the interstitial gas in the emulsion is limited primarily by the time it takes the xenon atoms to traverse the interstitial space, which is equivalent to the diffusion time. For a typical interstitial length $L = 100$ µm and gas diffusion $D = 3$ mm$^2 \cdot$s$^{-1}$, the emulsion-adsorbed phase exchange rate would be on the order of 1000 s$^{-1}$. Since the bubble-emulsion and emulsion-adsorbed phase gas exchange processes occur on different time scales, they can be treated independently in the NMR measurements.

The flow terms in Eq. 3 can be made negligible with the appropriate choice of NMR pulse sequence parameters. Note that the magnetization in each phase is uniform along the vertical bed direction if there is no net inflow or outflow of polarized gas during the gas exchange measurement, the gas is in a steady-state flow, and all $^{129}$Xe $T_1$ times are long compared to the flow process. The details of the pulse sequence timing and phase-cycling methods implemented to negate flow effects are described in Section 3 below, while confirmatory measurements of the very long $^{129}$Xe $T_1$ times in the adsorbed and emulsion phases for our sample are given in Section 4.

The absence of the gas inflow and outflow terms and the negligible $^{129}$Xe longitudinal relaxation rates lead to simplified equations for the dynamics of the net $^{129}$Xe magnetization in the three phases:

$$\frac{dM_b}{dt} = -K_{be}\left(M_b - \frac{\psi_b}{\psi_e}M_e\right)$$
$$\frac{dM_e}{dt} = -K_{be}\left(\frac{\psi_b}{\psi_e}M_e - M_b\right) - K_{ea}\left(M_e - \frac{\psi_e}{\psi_a}M_a\right). \quad (4)$$
$$\frac{dM_a}{dt} = -K_{ea}\left(\frac{\psi_e}{\psi_a}M_a - M_e\right)$$

To solve Eq. 4 for bubble magnetization, we assume time-independent volume fractions $\psi_b$, $\psi_e$ and $\psi_a$. In addition, with the appropriate choice of NMR pulse sequence (described below), we can destroy the initial emulsion and adsorbed phase magnetization, so that the initial conditions become:

$$M_b(0) = M_b^0,\ M_e(0) = M_a(0) = 0. \quad (5)$$

If the emulsion-adsorbed phase exchange rate is much faster than bubble-emulsion exchange rate, and if

$$K_{ea}\left(1 + \frac{\psi_e}{\psi_a}\right) \gg K_{be}\left(1 + \frac{\psi_b}{\psi_e}\right), \quad (6)$$

then the bubble magnetization is:

$$M_b(t) = \left[\frac{\psi_e + \psi_a}{\psi_b + \psi_e + \psi_a}\exp\left(-K_{be}\left(1 + \frac{\psi_b}{\psi_e + \psi_a}\right)t\right) + \frac{\psi_b}{\psi_b + \psi_e + \psi_a}\right]M_b^0. \quad (7)$$

Emulsion-adsorbed phase exchange may be treated similarly to bubble-emulsion exchange. By assuming the following initial conditions:

$$M_b(0) = 0,\ M_e(0) = 0,\ M_a(0) = M_a^0, \quad (8)$$

and applying the inequality in Eq. 6, the adsorbed phase exchange rate can be expressed as:

$$M_a(t) = \left[\frac{\psi_e}{\psi_a + \psi_e}\exp\left(-K_{ea}\left(1 + \frac{\psi_e}{\psi_a}\right)t\right) + \frac{\psi_e}{\psi_a + \psi_e}\right]M_a^0. \quad (9)$$

With the proper initial conditions (Eqs. 5 and 8), we are able to determine the exponential decay rates of the bubble and adsorbed phase magnetization, from which the exchange rates $K_{be}$ and $K_{ea}$ may be calculated by fitting the data to Eqs. 7 and 9.

## 3    Experimental

*3.1    Apparatus*

Figure 2 shows a schematic of the experimental apparatus.  A xenon hyperpolarization system provides polarized gas for the fluidized bed column, which is located in the bore of the NMR magnet.  The gas flow from the laser-polarization apparatus, through the fluidizing column, is maintained by a vacuum pump and a mass flow controller placed downstream of the magnet.

$^{129}$Xe is spin-polarized using the spin-exchange optical pumping method [32].  Spin exchange occurs in a ~ 80 cm$^3$ glass cell that contains a small amount of rubidium metal, the intermediary for the spin exchange process. The total gas pressure in the optical pumping cell is maintained at 2.5 bar, with the mixture consisting of ~ 95% xenon (26.4% $^{129}$Xe) and ~ 5% N$_2$.  We heated the cell to 130ºC to create an appropriate Rb vapor density for spin polarization of the Rb atoms via optical pumping on the Rb D1 line (~ 795 nm).  Circularly-polarized light for optical pumping is supplied by a 30 W broad-spectrum (~ 2 nm) FAP Laser Diode Array [Coherent Systems Inc., Santa Clara].  After a few minutes of optical pumping, Rb-Xe collisions boost the $^{129}$Xe spin polarization to ~ 0.5 - 1%.  The polarizer operates in continuous flow mode from the supply bottle to the fluidized bed column, with $^{129}$Xe becoming spin polarized as it flows through the optical pumping cell.

The polarized xenon flows from the polarizer to the fluidized bed column, and then onto the vacuum pump, through 1/8" i.d. Teflon tubing.  The pressure potential is supplied by the pump located at the end of flow path.  The mass-flow controller, placed just before the vacuum pump, mediates the influence of the pump and provides a steady, continuous gas flow at rates ranging from 20 to 200 standard cm$^3$·min$^{-1}$ (sccm). We measured the xenon gas pressure in the particle bed to be ~ 2.5 bar, (the pressure drop across the bed was negligible).  In our experiments, the gas flow and particle motions were in a steady state, and we minimized transient and unstable bed behavior with a steady, homogeneous gas flow from the bottom of the column.

The fluidized bed column, integrated with a vertical bore NMR probe body, is shown in Figure 3.  The fluidizing bed apparatus consists of a 6 mm i.d. cylindrical Pyrex column, a windbox below the column, and two gas diffusers made of porous polyethylene [Porex Corp., Atlanta GA] with a pore size of ~ 20 µm. The windbox, with its large volume, provides a buffer for the gas flow from the polarizer and the column. On top of the windbox is a gas diffuser, which ensures the vertical gas flow is homogeneous in cross-section. The column, located above the diffuser, holds the bed of particles, while a second diffuser covering the top of the column prevents particles from leaving the bed. The entire apparatus is assembled from non-metallic materials, in order to minimize spin depolarization of the hyperpolarized $^{129}$Xe during transport, and to ensure compatibility with the NMR environment.

The probe body with the fluidized bed and associated gas-transport tubing was positioned in a 4.7 T vertical wide-bore magnet [Oxford Instruments, Oxford, UK], which is interfaced to an AMX2-based NMR console [Bruker Biospin, Billerica, MA]. We conducted $^{129}$Xe NMR experiments at a frequency of 55.4 MHz using a homebuilt Alderman-Grant style RF coil [36]. The coil was constructed from copper foil glued onto a 1/2" o.d. Teflon tube. The sample was positioned in the homogeneous $B_1$ region of the coil, with the inner radius of the fluidized bed column less than 50% of the coil radius. The total sensitive length of the coil is ~ 2.4 cm, which covers a section in the middle of the particle bed. This configuration eliminates possible artifacts in the NMR spectra due to the inclusion of bulk xenon outside the bed appearing in the sensitive region of the coil [13].

Aluminum oxide ($Al_2O_3$) powder [EMD Chemicals, Gibbstown, NJ] with an average density of 3.97 g·cm$^{-3}$ was used as the solid phase of the fluidized bed. We sieved the particles to create a sample with a particle size distribution ~ 75 - 106 µm in diameter, so that according to the Geldart classification scheme, they belong to Geldart Group B. The shape of the $Al_2O_3$ particles is highly irregular. This irregularity, along with a large magnetic susceptibility of diamagnetic aluminum oxide ($-37 \times 10^6$ cm$^3$·mol$^{-1}$) [37], increases the linewidth of $^{129}$Xe gas flowing through the interstitial spaces between the particles ($\Delta \nu$ was approximately 650 Hz vs. 75 Hz for bulk gas). In addition, the surface of $Al_2O_3$ particles is not smooth, but contains micropores of size ~ 5 µm. These pores significantly increase the surface-area-to-volume ratio of the particles, so that the contact area between the gas and the solid phase is very large. (A standard BET surface-area measurement [38, 39] on a representative sample of the $Al_2O_3$ particles yielded a surface-area of 99.8 m$^2$·g$^{-1}$.) When moving close to the surface of a particle, $^{129}$Xe atoms may temporarily adsorb on the surface [40, 41] resulting in a chemical shift of $^{129}$Xe resonant frequency.

To estimate the relaxation terms in Eq. 3, we performed a $^{129}$Xe $T_1$ measurement on a sealed glass cell filled with $Al_2O_3$ particles and containing ~ 3 bar of ~ 95% enriched $^{129}$Xe. The glass cell had a volume of ~ 200 cm$^3$, and was placed in a ~ 6 cm i.d. Alderman-Grant coil [Nova Medical, Wilmington, MA] tuned to 55.4 MHz. We conducted the experiment using a horizontal bore 4.7 T MRI magnet interfaced to an Avance NMR console [Bruker Biospin, Billerica, MA]. The sample was sealed and did not employ the laser-polarization apparatus - instead, the $^{129}$Xe spins reached thermal equilibrium. Since the emulsion and adsorbed phase peaks were well resolved, we could measure the $T_1$ of both phases simultaneously using the saturation recovery technique [17].

*3.2    NMR Measurement of Bubble-Emulsion Exchange*

Magnetic susceptibility differences between the particles and the xenon gas, and resultant field gradients at the solid – gas interfaces, result in $^{129}$Xe spins in the emulsion phase experiencing significantly larger field heterogeneity than spins in the bubbles. Additionally, the effective field inside the bubble is more homogeneous, due to the bubble's spherical shape, than in the emulsion phase. As a result, the NMR spectral line from the bubble-phase $^{129}$Xe is significantly narrower than that from the emulsion phase (See Figure 4), and therefore, a sequence employing $T_2$ contrast can filter the emulsion and adsorbed phase peaks from the spectrum, allowing the bubble phase signal to be clearly resolved and measured without spectral overlap.

We employed a longitudinal *z*-storage 3-pulse sequence, based on traditional NMR exchange spectroscopy measurements [16,42], to generate $T_2$ contrast and measure the bubble-emulsion exchange rate. A $\pi/2$-$\pi$-$\pi/2$ pulse train was used to generate $T_2$ contrast. The spin-echo time, *TE*, was set longer than the $^{129}$Xe $T_2$ in the emulsion and adsorbed phases (~ 2 ms), such that the second $\pi/2$ pulse stored only the bubble magnetization along the longitudinal axis. Gas exchange between the bubble and emulsion phases occurred during the subsequent exchange time, $\tau$, which was varied from 0.2 – 30 ms. The third $\pi/2$ pulse sampled the remaining bubble magnetization.

We took a number of precautions to ensure the bubble exchange measurement was not impacted by either the inflow or outflow of polarized $^{129}$Xe spins during the exchange period. A preparatory, non-selective $\pi/2$ saturation pulse was applied to destroy the gas magnetization within the RF coil, after which freshly polarized gas filled the bottom portion of the RF coil during a carefully determined delay time, based on the flow rate. This preparation ensured the exchange measurement was performed only on spins that were initially in this bottom portion of the coil, and that RF-tagged spins did not leave the RF coil during the exchange sequence. To ensure signal from inflowing bubbles during the exchange time was not convolved with the exchange signal due to the third $\pi/2$ pulse, a simple phase-cycling scheme was implemented. The first $\pi/2$ pulse and receiver phase were alternated by $\pi$ on every second scan while the third $\pi/2$ pulse was unchanged from scan to scan. This method cancelled the inflowing bubble magnetization (which is only influenced by the readout pulse) after every second scan, while the original magnetization was added from scan to scan as it was cycled with the receiver phase. Finally, we note that it was not necessary, for the accuracy of the measurement, to exclude inflowing emulsion or adsorbed-phase xenon, because the upward flowing, RF-tagged bubbles are not in contact with the inflowing emulsion gas. However, to avoid the need to deconvolve the resultant spectrum into the bubble-phase-only signal, we acquired the FID signal from the third π/2 pulse after a delay time of 1 ms. This delay eliminated the NMR signal from polarized $^{129}$Xe spins that had entered the emulsion and adsorbed phases within the RF coil during the exchange period, as well as emulsion or adsorbed-phase signal resulting from polarized $^{129}$Xe spins that had exchanged out of bubble during the exchange measurement, leaving only signal due to $^{129}$Xe spins originally in bubble that had not transferred out during the exchange time.

By plotting the bubble signal as a function of exchange time, $\tau$, we were able to determine the bubble signal decay rate, $R_b$, which is related to the bubble-emulsion exchange rate, $K_{be}$, according to Eq. 7:

$$R_b = K_{be}\left(1 + \frac{\psi_b}{\psi_a + \psi_e}\right). \tag{10}$$

*3.3  NMR Measurement of Emulsion-Adsorbed Phase Exchange*

Aluminum oxide powders have a micro-porous surface that promotes xenon adsorption, and leads to a large $^{129}$Xe chemical shift. This feature of $Al_2O_3$ allows us to distinguish between the adsorbed and free-gas phases and therefore to measure the gas exchange between the emulsion and adsorbed phases. The emulsion-adsorbed phase exchange can be measured if the bubble-emulsion exchange is very slow by comparison, and therefore negligible. The rate-limiting step for emulsion-adsorbed phase exchange is gas diffusion in the interstitial spaces of the emulsion – we estimated this time scale to be ~ 1 ms.

To measure the emulsion-adsorbed phase exchange rate, we used a pulse sequence based on the standard saturation-recovery experiment. A π/2 selective saturation pulse was used to destroy the adsorbed phase magnetization in the RF coil region of the column. After the saturation, a delay time, $\tau$ (varied from 5 µs to 5 ms), was used to allow exchange to occur between the adsorbed and emulsion phases. Spins exchanging in from the emulsion phase retained their high spin polarization. Finally, a non-selective π/2 RF pulse allowed instantaneous sampling of the $^{129}$Xe magnetization in all three phases after exchange.

Due to RF power limitations, the duration of the selective RF pulse was limited to > 1 ms, which is of the same order as the expected emulsion-adsorbed phase exchange time. As a result, the saturation of the adsorbed phase magnetization was not complete. To optimize the destruction of the adsorbed phase magnetization with respect to the bubble and emulsion signals using a 1 ms RF pulse, we offset the RF pulse such that there was a maximum decrease in the RF pulse excitation profile between the adsorbed phase and emulsion/bubble phase peaks. This technique enabled us to reduce the $^{129}$Xe adsorbed phase peak magnitude by 50% from its equilibrium value, which sufficed for an accurate measurement of emulsion-adsorbed phase exchange rate. Precautions against inflow and outflow effects in this experiment were much less relevant than in the bubble-emulsion exchange measurement. The particles, and consequently the adsorbed $^{129}$Xe phase have no net upward velocity, so outflow effects are negligible. Additionally, due to the very short timescale of the measurement, the amount of freshly polarized gas flowing in or out of the RF coil sensitive region approximated ~ 0.2% of its volume.

The adsorbed phase $^{129}$Xe recovery rate, $R_a$, was obtained by fitting an exponential function to the adsorbed phase signal, after deconvolution, to obtain accurate $^{129}$Xe adsorbed phase integrals. The emulsion-adsorbed phase exchange rate, $K_{ea}$, could then be determined from $R_a$ according to Eq. 9:

$$R_a = K_{ea}\left(1 + \frac{\psi_e}{\psi_a}\right), \tag{11}$$

where $\psi_e/\psi_a = 2.01 \pm 0.16$, independent of flow rate, from our experiments above.

## 4  Results

We measured the $T_1$ times for $^{129}$Xe in the emulsion and adsorbed phases using a sealed glass cell filled with $Al_2O_3$ particles and enriched $^{129}$Xe gas, as described above. The observed $^{129}$Xe $T_1$ relaxation time was the same in the two phases ($T_1^{obs}$ = 44.0 s) due to the fast exchange between them. The observed relaxation time is related to the intrinsic $^{129}$Xe $T_1$ in the emulsion, $T_1^e$, and adsorbed phase, $T_1^a$, by

$$\frac{1}{T_1^{obs}} = \frac{\psi_e}{T_1^e} + \frac{\psi_a}{T_1^a}, \tag{12}$$

where $\psi_e + \psi_a = 1$, in the absence of bubbles. It follows from this relationship that $T_1^e > \psi_e T_1^{obs}$ and $T_1^a > \psi_a T_1^{obs}$. The value of $\psi_e$ was measured to be 0.66 ± 0.03, in this case. The lower limits for $T_1$ in the emulsion and adsorbed phases were found to be: $T_1^e > 29$ s and $T_1^a > 14.5$ s. Finally, while we could not measure the $T_1$ relaxation time for $^{129}$Xe inside the bubble, $T_1^b$, we expect it does not differ significantly from the relaxation time of bulk xenon, ~ 1000 s. Compared to the estimated exchange rates $K_{be}$ ~ 100 s$^{-1}$ and $K_{ea}$ ~ 1,000 s$^{-1}$, the relaxation rates for $^{129}$Xe in the three phases are much smaller and thus the $T_1$ terms in Eq. 3 may be neglected. A summary of the $^{129}$Xe relaxation times, estimated exchange rates and timescales of flow processes in our fluidized-bed experiments, as well as timescales for the different NMR measurements are provided in Table 1.

Figure 4 shows $^{129}$Xe spectra from the $Al_2O_3$ particle bed, acquired at 13 different flow rates ranging from 20 – 180 sccm. The broad peak (~ 650 Hz FWHM) at lower frequency is due to xenon gas in the emulsion phase. Its broad line-width results from the large field gradients experienced in the interstitial spaces of the emulsion. There is a slight shift in frequency of this peak away from free xenon gas. A second broad peak with roughly the same width, but shifted 2.3 kHz from the emulsion peak, was identified as the adsorbed $^{129}$Xe gas phase. The third peak (~ 75 Hz FWHM) on the shoulder of the emulsion peak grows with increasing flow rate. Since the number and size of bubbles increase with flow rate [20], we identified this peak as originating from the xenon bubbles.

Although the emulsion and adsorbed phase peaks are almost completely resolved, the emulsion and bubble signals overlap completely [13]. To accurately determine the volume fractions, $\psi_p$, of each phase $p$, we fit the $^{129}$Xe spectra (such as those in Figure 4) to three complex Lorentzian functions of the form:

$$\frac{a}{b + i(\omega - c)}. \qquad (13)$$

This enabled us to deconvolve the three peaks and compute the area under each peak.

Figure 5 shows the expansion of the bed, and the NMR-derived fractional volume, $\psi_p$, of xenon in the bubble, emulsion and adsorbed phases for flow rates between 20 and 180 sccm. The bed expansion was determined from visual measurements of bed height, and is due to the onset of homogeneous fluidization, below ~ 40 sccm, and then the effect of bubbles appearing in the bed, consistent with the spectra of Figure 4. As in Figure 4, the bubble volume in Figure 5b increases with flow rate. In addition, Figure 5b shows that the fractional emulsion and adsorbed phase volumes decrease with flow rate. This observation is consistent with the increase in bubble volume and the steady-flow condition. As the bubbles increase in volume and number, they displace the solid particles of the bed and their interstitial gas from the region of the column within the RF coil, thus reducing fractional emulsion- and adsorbed-phase volumes by the same amount. The ratio, $\psi_e/\psi_a$, remains constant at $2.01 \pm 0.16$ as a result.

To understand the variations in bubble size, we acquired 256 systematic measurements of the bubble signal over an extended time period. Figure 6 shows the bubble signal (~ volume) distribution for two different bed configurations (both obtained at flow rate of 80 sccm). The narrower distribution on the left was obtained when the diffuser was positioned right below the coil sensitive region, while the distribution on the right was obtained when the diffuser was placed 0.25 cm below the coil. When the diffuser is positioned lower in the fluidized bed column, the bubbles have more time to grow and coalesce, resulting in a wider bubble size distribution. The change in placement of the diffuser corresponds to only about 10% of the coil length, resulting in only a moderate improvement in the bubble size distribution at the higher position for simple spectra acquired from the entire RF coil region. However, the higher diffuser placement greatly enhanced the reproducibility of the bubble-emulsion exchange experiments as these are only performed on xenon in the lower portion of the RF coil, due to the pre-saturation employed to prevent xenon outflow during the measurement, as described in Section 3.2.

The xenon bubble signal decay rate, $R_b$, was obtained from the area of the $^{129}$Xe bubble peak as a function of exchange time from 1 – 30 ms. The time-dependence of the bubble magnetization acquired at a flow rate of 80 sccm, as well as an exponential fit to the data, are plotted in Figure 7a. The experimental data can be modeled by a single-exponential, indicating the dominance of a single exchange process. The xenon bubble-emulsion exchange rate, $K_{be}$, for this flow rate was then calculated according to Eq. 10. Values of $K_{be}$ measured at a few representative flow rates are given in Table 2.

Similarly, the adsorbed phase recovery rate, $R_a$, was derived from the area of the deconvolved $^{129}$Xe adsorbed phase peak as a function of exchange time from 5 µs – 5 ms. The time-dependence of the adsorbed phase signal acquired at a flow rate of 60 sccm, and an exponential fit to the data, are plotted in Figure 7b. The xenon emulsion-adsorbed phase exchange rate, $K_{ea}$, for this flow rate was then calculated according to Eq. 11. Values of $K_{ea}$ measured at a few representative flow rates are given in Table 2.

## 5   Discussion

As reported in Table 2, $K_{be}$ shows a slight decrease with flow rate, while $K_{ea}$ does not vary, within the bounds of uncertainty, which is consistent with well-known models of gas-fluidized beds [4]. For Geldart Group B particles, the expansion of the emulsion phase after the onset of fluidization is negligible, and consequently the inter-particle distance in the emulsion should be independent of flow rate. Since $K_{ea}$ depends primarily on the ability of gas molecules to diffuse within the interstitial space, it should be independent of flow rate in agreement with our experimental results. A goal of future work is a systematic study of gas exchange rate variation with flow rate.

The major contributor to the uncertainty in $K_{be}$, as reported in Table 2, are statistical variations in bubble size distribution. The generation and coalescence of bubbles are uncontrollable random processes, resulting in variations in the number and size of bubbles in the particle bed at any instant of the measurement. Since these variations are larger at low flow rates, when the number of bubbles is small, the measurements at low flow rates have a greater uncertainty than those at higher flow rates. There is also an additional systematic error in the estimation of $K_{be}$, which is due to the bubble size being non-uniform along the column (bubbles increase in size as they rise up the bed). As a result, the average bubble volume fraction, $\psi_b$, is overestimated for short exchange times and underestimated for long exchange times.

With sufficient SNR to allow measurements to be made without signal averaging, we could monitor the bubble volume in each shot by acquiring the signal from the first π/2 pulse of the exchange measurement. This bubble volume measure would allow for subsequent "binning" of the resulting data, acquired with varying exchange time, τ, based on the representative bubble signal measurement. A resultant $K_{be}$ value could then be determined for each bubble volume or range of volumes. With hyperpolarized noble gases, it should be possible to provide the fluidizing gas with sufficient levels of polarization to remove the necessity to signal average. We are currently working to improve gas polarization and delivery for our fluidized bed apparatus, with the aim of increasing polarization delivered to the column by a factor of ~ 20 – 50. Our efforts in this area include: improved delivery of gas to the column and a reduction of polarization loss in the tubing; different operating modes for the polarizer to increase polarization levels obtained in the optical-pumping cell; implementation of novel laser technology to increase resonant laser power by a factor of ~ 5 – 10 [43, 44]; and using $^3$He as the fluidizing gas. $^3$He generally yields a ~ tenfold increase in NMR signal over $^{129}$Xe for a given polarization level [45]. We are confident that a combination of some of the approaches above will yield the required level of signal to allow for single-scan bubble-exchange measurements, which can then be performed as a function of bubble volume variation in the RF coil, or in a spatially-resolved manner after the incorporation of NMR imaging methodologies into our exchange sequences.

As can be seen in Table 2, uncertainties in $K_{ea}$ are much lower than for $K_{be}$. Inflow and outflow effects are negligible in this case, as the solid particles are essentially static in the column. Additionally, the time-scale for emulsion-adsorbed phase exchange is so much shorter than for the bubble-phase measurements, making gas inflow effects negligible. Perhaps the largest contributor to this measurement is the incomplete RF saturation of the adsorbed phase signal, due to power limitations of the RF pulse and the rate of exchange onto the particles which is of the same order as the pulse length.

## 6  Conclusions

We have applied NMR techniques to non-invasively measure gas exchange rates between the three gas phases in a gas-fluidized particle bed; specifically, bubble-emulsion and emulsion-adsorbed phase exchange in a bed of $Al_2O_3$ particles fluidized by hyperpolarized xenon gas. The contrasting values of the transverse relaxation time of gas atoms in the bubble and particle-dense phases allowed us to differentiate the phases effectively using a $T_2$-weighted variation of the traditional NMR exchange spectroscopy pulse sequence. In addition, the chemical shift of xenon adsorbed on the surface of the fluidized particles allowed us to measure the emulsion-adsorbed phase exchange rate with a simple saturation-recovery method. The ability to measure both bubble-emulsion and emulsion-adsorbed phase exchange simultaneously in a three-dimensional system, is a feature of the NMR method that is not possible with any other current technique.

Potential future studies include measuring the exchange rates in a spatially-resolved manner, or the investigation of gas flow dynamics in the homogeneous fluidization regime, where debates exist regarding whether or not the particles are completely suspended by the fluidizing gas and behave like a fluid. Pulsed field gradient NMR methods will also be used to probe the coherent and dispersive motions of the gas phase. Finally, the use of NMR-detectable particles opens the potential for us to probe both gas and particle phases together in a single system.

## 7 Acknowledgements

We acknowledge support for this work from NSF grant CTS-0310006, NASA grant NAG9-1489, Harvard University and the Smithsonian Institution.

| Process | Time Scale |
|---|---|
| $^{129}$Xe $T_1$ relaxation in bubbles | > 10 min |
| $^{129}$Xe $T_1$ relaxation in the emulsion | > 30 s |
| $^{129}$Xe $T_1$ relaxation in the adsorbed phase | > 15 s |
| Bubble flow through the coil region (50-160 sccm) | 400 – 200 ms |
| Emulsion gas flow through the coil region (constant after bed is fluidized) | 4.8 s |
| Time per scan in the bubble-emulsion exchange measurement | 100 ms |
| Overall bubble-cloud exchange (estimated) | 12 ms |
| Overall cloud-emulsion exchange (estimated) | 1 ms |
| Overall bubble-emulsion exchange (estimated) | 12 ms |
| Time per scan in the emulsion-adsorbed phase exchange measurement | 10 ms |
| Emulsion-adsorbed phase exchange (estimated) | 1 ms |

**Table 1**: Time-scales of physical processes and NMR parameters in the fluidized-bed experiments. Details of the bubble-cloud and cloud-emulsion exchange estimates are based on the models in [4], but are beyond the scope of this paper.

| Flow rate (sccm) | $K_{be}$ (s$^{-1}$) | Flow rate (sccm) | $K_{ea}$ (s$^{-1}$) |
|---|---|---|---|
| 80 | 236 ± 44 | 50 | 1386 ± 31 |
| 100 | 221 ± 43 | 60 | 1367 ± 29 |
| 140 | 173 ± 34 | 70 | 1312 ± 29 |

**Table 2**: Example bubble-emulsion (left) and emulsion-adsorbed phase (right) exchange rates (in units of s$^{-1}$) at different flow rates (standard cubic centimeters per minute).

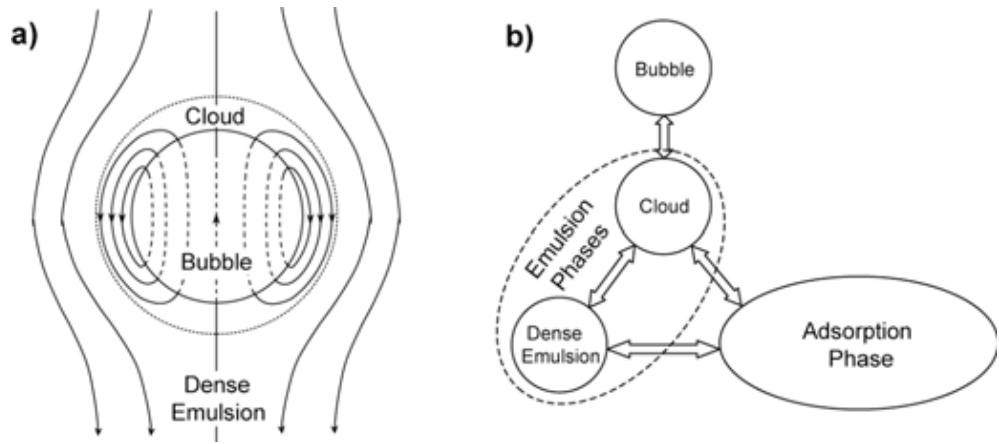

**Figure 1: a.** Gas-flow streamlines through and around an idealized spherical bubble in a gas-fluidized bed, as seen from the bubble reference frame. In this study, we consider the cloud sub-phase as part of the overall emulsion phase. **b.** Gas-exchange pathways between the gas phases present in a bubbling fluidized bed.

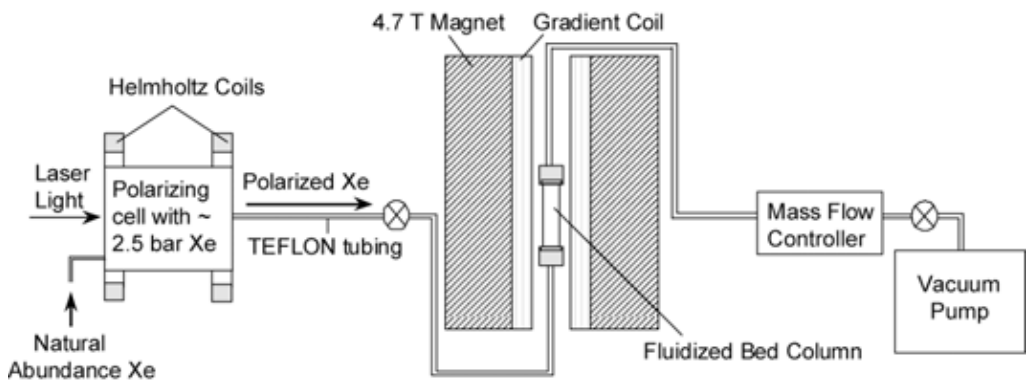

**Figure 2**: A schematic diagram of the hyperpolarized xenon - fluidized bed - NMR apparatus. Narrow 1/8 inch i.d. Teflon tubing connects the different sections of the apparatus and provides the gas flow path. The mass flow controller moderates the effect of the pump and determines the gas flow in the particle bed.

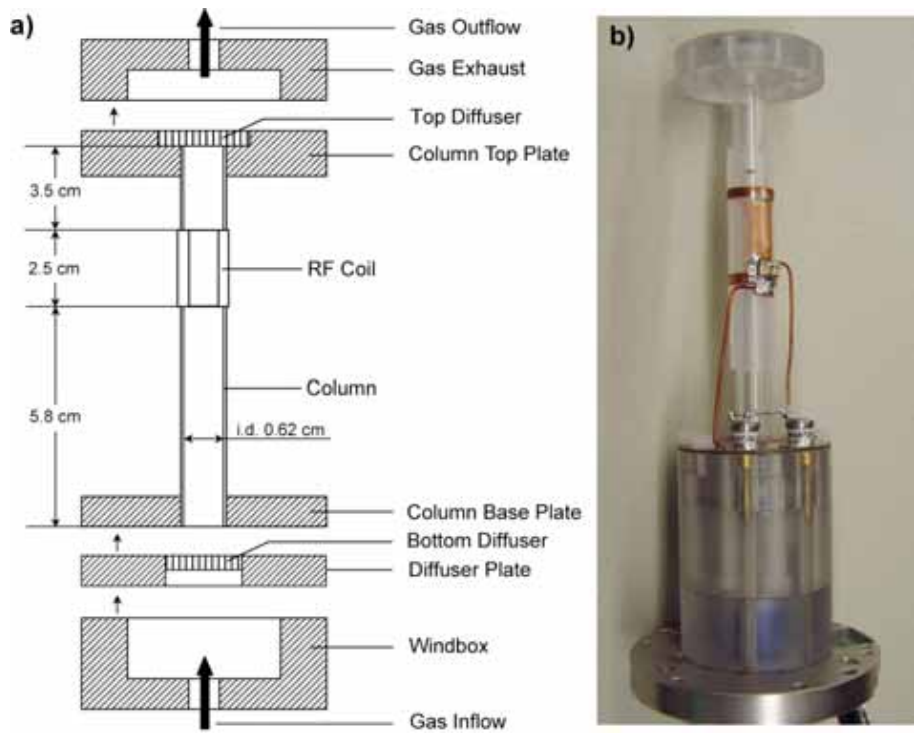

**Figure 3: a.** Schematic of the fluidized bed column and its components. **b.** Photograph of the fluidized bed apparatus integrated on the vertical-bore NMR probe body and including the RF coil,

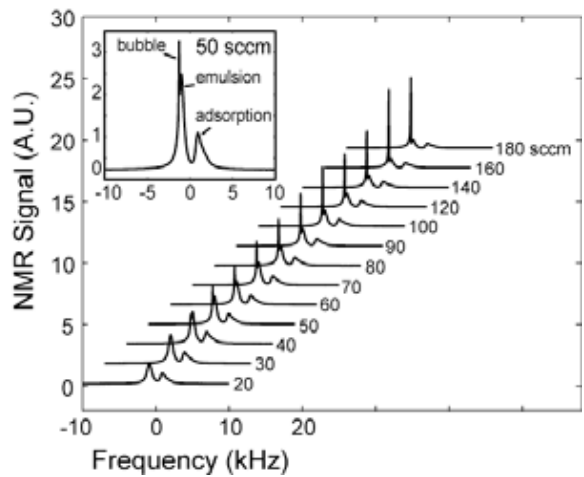

**Figure 4**: $^{129}$Xe gas spectra measured in an $Al_2O_3$ particle bed at flow rates ranging from 20 to 180 sccm. The inset shows an individual spectrum obtained at 50 sccm. The three peaks in the spectra correspond to xenon in the bubble, emulsion and adsorbed phases. The spectra were acquired without any presaturation,

using π/2 pulses exciting the entire RF coil region, 2 signal averaging scans and a repetition rate, *TR*, of 9.2 – 1.5 s (*TR* decreases with increasing flow rate), across a spectral width of 20 kHz.

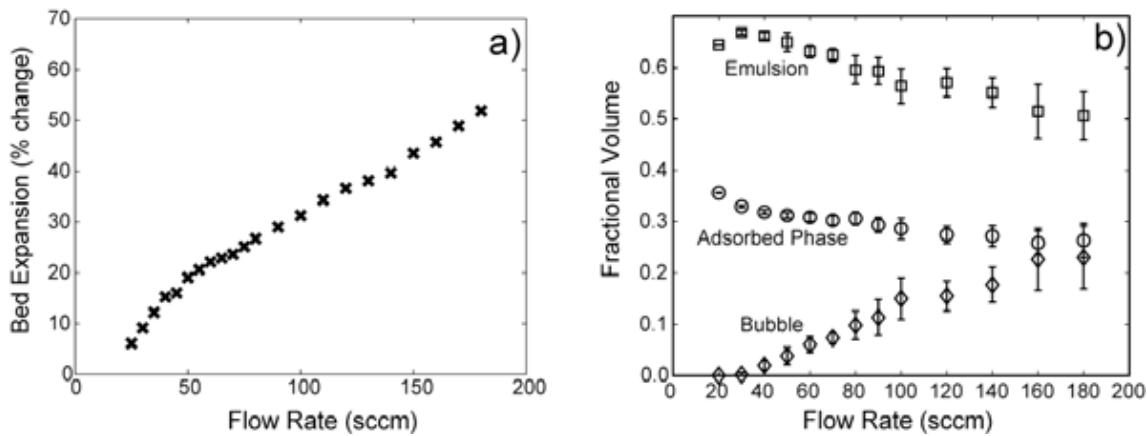

**Figure 5**: a) Expansion of the alumina particle bed, from visual measurements of particle bed height. Initial expansion is due to fluidization, above ~ 40 sccm the expansion is due to bubble formation. b) Fractional volumes of $^{129}$Xe within the bubble (◊), emulsion ( ) and adsorbed (o) phases as a function of flow rate. Each point is the average of 32 separate single-shot spectra acquired over a 5 min period, using non-spatially selective π/2 pulses, with a delay between spectra acquisitions of 9.2 s.

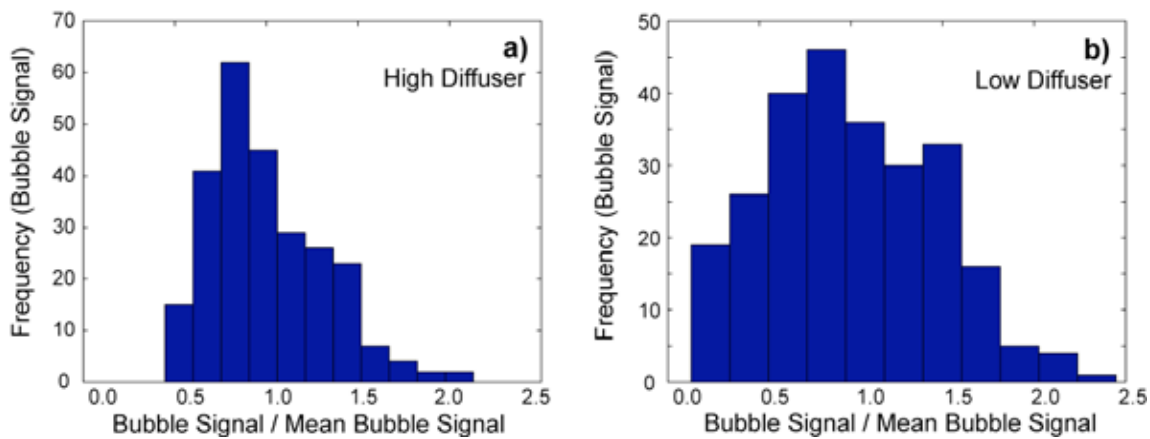

**Figure 6**: Bubble volume distribution for 256 separate $^{129}$Xe spectra acquired from the fluidized bed with the diffuser positioned just below (left) and 0.25 cm below (right) the RF coil-sensitive region. The variation in bubble volume is larger when the diffuser is placed lower in the fluidized bed, since the bubble volume begins to change randomly as bubbles rise up the bed. Both volume distributions were obtained at a flow rate of 80 sccm. Each of the 256 spectra were acquired using non-spatially selective π/2 pulses, 1 signal averaging scan and a repetition rate of 2.5 s, across a spectral width of 20 kHz.

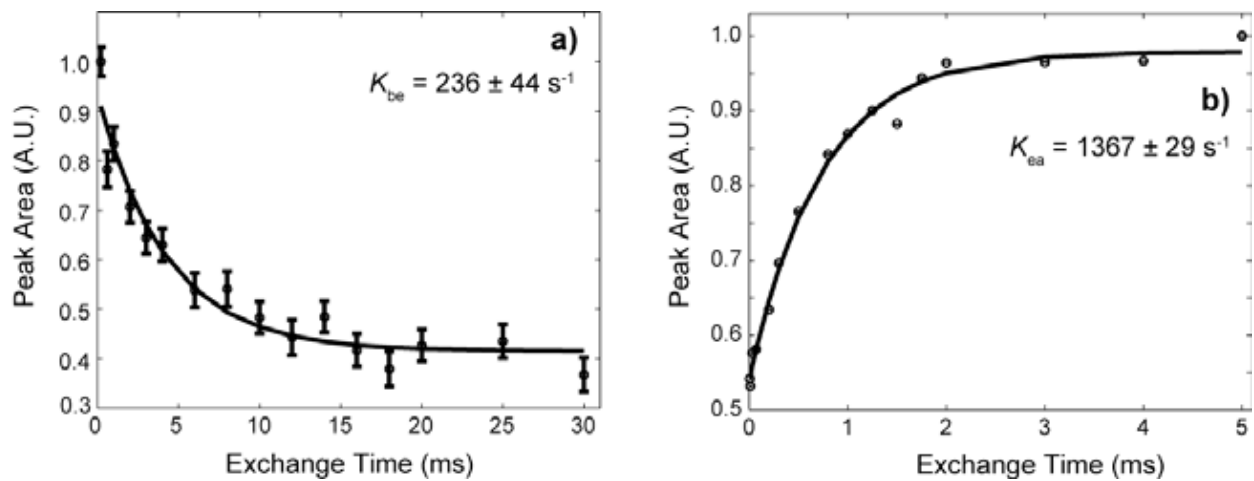

**Figure 7**: **a**. Change in bubble signal intensity for exchange times, $\tau$, varying from 1 – 30 ms, measured at a flow rate of 80 sccm, using the $T_2$-contrast exchange spectroscopy sequence described in Section 3.2. For each $\tau$, 250 signal averaging scans were acquired. The delay between pre-saturation and the exchange pulse sequence for this flow rate was 75 ms. The signal decreases due to the magnetization transfer from the bubble into the emulsion phase. **b**. Change in adsorbed phase signal intensity as a function of exchange times varying from 0.005 – 5 ms, measured at a flow rate of 60 sccm, using the saturation recovery-based sequence described in Section 3.3. For each $\tau$, 4 signal averaging scans were acquired at a repetition rate of 4 s, The signal increases because of the magnetization transfer from the emulsion into the adsorbed phase.